\documentclass[12pt]{article}
\usepackage{graphicx}

\begin{document}
\begin{titlepage}
%
%






{\ }









\begin{center}
\centerline{Heating in collisions of solids: possible application to impact craters}
\end{center}
\begin{center}
\centerline{V.Celebonovic}			
\end{center}

\begin{center}
\centerline{Institute of Physics,Pregrevica 118,11080 Zemun-Belgrade,Serbia}
\centerline{vladan@ipb.ac.rs}
\end{center}





\begin{abstract}
Due to the importance of collisions and impacts in early phases of the evolution of the planetary system, it is interesting to estimate the heating of a solid target due to an impact in it . A physically simple calculation of the temperature to which  a solid target heats up after the impact of a projectile with mass $m$ and speed $v$ is performed,and possibilities for the application of this result in planetology are pointed out.\footnote{to appear in Serbian Astron.Journal}
\end{abstract}
\end{titlepage}


{


\section{INTRODUCTION}

Colliding macroscopic bodies exchange momentum. In inellastic collisions,a part of the momentum is transformed into heat. 

The aim of this paper is to analyze to some extent the thermal effects in a collision of solid bodies of different masses, and to estimate the temperature to which the more massive of the colliding bodies (the target) heats up. If the temperature of the target rises sufficiently, melting and ultimately vaporisation will occur. The melting temperature of a solid can be estimated by the so called Lindemann criterion. The value obtained by the Lindemann criterion will be compared with  the temperature to which the target heats up in a collision. As the temperature to which the target heats up depends on the kinetic energy of the impactor this calculation will provide an estimate whether or not in an impact the target melts,and as a consequence can it be analyzed by applying solid state physics or not. 

If the mass of the target is sufficiently higher than the mass of the impactor, the impact will produce virtually no change in the momentum of the target, while the impactor will almost certainly be destroyed. The kinetic energy of the impactor will be spent on heating of the target and on introducing changes into its structure.

The following section contains the calculation of the temperature to which a target in a collision heats up in the case of an inellastic collision. The calculation is performed for two cases: for the "ideal" case, in which all of the kinetic energy of the impactor is converted into heat,and for a more realistic case,in which a part of the kinetic energy is used to form a crack in the target. The third part shows how to include the equation of state of the material of the target in the calculation in one particular example,and the last part of the paper points to possible planetological application of the calculation presented in the present paper.    


\section{CALCULATIONS}

\subsection{The ideal case}

The specific heat of a solid at low temperature is given by ([3]) 
\begin{equation}
C_{V} = \frac{2 \pi^{2}}{5} \frac{k_{B}}{(\hbar \overline{V})^{3}}(k_{B}T)^{3}
\end{equation}
where $k_{B}$ is Boltzmann's constant, $\hbar$ is Planck's constant divided by $2\pi$,$T$ the absolute temperature and $\overline{V}$ the speed of sound waves in the material. It can be shown in solid state physics that
\begin{equation}
\overline{V}= (\frac{\partial P}{\partial \rho})^{1/2}	
\end{equation}
where $P$ and $\rho$ denote,respectively, the pressure and density of the material. Inserting eq.(2) into eq.(1),it follows that 
\begin{equation}
	C_{V}= \frac{2 \pi^{2}}{5} \frac{k_{B}}{(\hbar)^{3}}(\frac{\partial P}{\partial \rho})^{- 3/2}(k_{B}T)^{3}
\end{equation}
Heating up volume $V$ of the material of the target by one degree requires the quantity of energy equal to $V C_{V}$. The temperature to which this volume will heat up when impacted by an object having mass $m$ and speed $v$ is in the ideal case given by
\begin{equation}
T_{0}+	\frac{1}{2}(\frac{m v^{2}}{V C_{V}})=T_{1}
\end{equation}
where $T_{0}$ denotes the intial temperature of the target.
Practically, the volume $V$ physically represents the volume of a crater resulting from the impact. The volume is obviously dependent on the shape of the crater. Assume that a crater has the shape of a half of a rotating ellipsoid with distinct semi axes,denoted by $a$,$b$ and $c$. Physically,$a$ and $b$ are the semi axes of the "opening" of the crater,and $c$ denotes its depth. The volume is $V=(2/3)\pi a b c$,which leads to the final expression for the temperature to which this volume of the target heats up as a result of the impact
\begin{equation}
T_{1}=T_{0}+ \frac{15}{8 \pi^{3} k_{B}}\frac{m v^{2}(\hbar \overline{V})^{3}}{a b c}(k_{B} T_{0})^{-3}  
\end{equation}
Inserting eq.(2) into eq.(5), it finally follows that
\begin{equation}
T_{1}=T_{0}+ \frac{15}{8 \pi^{3} k_{B}}\frac{m v^{2}\hbar^{3} (\frac{\partial P}{\partial \rho})^{3/2} }{a b c}(k_{B} T_{0})^{-3}  	
\end{equation}

Is this heating sufficient or not to lead to melting and even vaporisation of the material of the target? This can be detemined by applying the so called Lindemann criterion which states that ([2])
\begin{equation}
	T_{m}= T_{m0} (\frac{\rho_{0}}{\rho})^{2/3}\exp(2\gamma_{0}(1-(\rho_{0}/\rho))) 
\end{equation}
In this expression $T_{m}$ denotes the melting temperature of a material at mass density $\rho$, $T_{m0}$ is the melting temperature at density $\rho_{0}$.The symbol $\gamma_{0}$ denotes the value of the Gruneisen parameter of the material at density $\rho_{0}$.The Gruneisen parameter is defined as 
\begin{equation}
	\gamma = \frac{\alpha K_{T}}{C_{V} \rho}
\end{equation}
where $\alpha$ is the thermal expansion coefficient and $K_{T}$ the isothermal bulk modulus of the material. Comparing eqs.(6) and (7) it follows that $T_{1}\leq T_{m}$ if
\begin{eqnarray}
\frac{E_{k}}{a b c}(\frac{\partial P}{\partial \rho})^{3/2} \hbar^{3}(k_{B} T_{0})^{-3}\leq\frac{4 \pi^{3}}{15 \hbar^{3}}T_{m0}(\frac{\rho_{0}}{\rho})^{2/3}\times \nonumber\\
\exp(2\gamma_{0}(1-(\rho_{0}/\rho)))\times[1-\frac{T_{0}}{T_{m0}}\nonumber\\\times(\frac{\rho_{0}}{\rho})^{-2/3}\exp(-2\gamma_{0}(1-(\rho_{0}/\rho))]
\end{eqnarray}
where $E_{k}$ denotes the kinetic energy of the impactor in the moment of impact. 

In practical terms,the fulfillement of this condition means that as a consequence of an impact,the target heats up at the point of impact, but does not melt. A further implication is that such an impact can be analyzed by solid state physics. The analytical expression of the derivative $\frac{\partial P}{\partial \rho}$ depends on the form of the equation of state of the material of the target.

\subsection{ The "real" case}

If the relative velocity of two colliding solid bodies is high enough,and if the material strength of the target is small enough,it can be expected that a fracture will occur in the target as a consequence of the collision. A standard result for the stress necessary for the occurence of a fracture is given by (for example,[7])
\begin{equation}
	\sigma_{C}=\frac{1}{2}(\frac{E \chi\tau}{r_{0}a})^{1/2}
\end{equation}
where $E$ is Young's modulus of the material,$\chi$ is the surface energy,$\tau$ the radius of curvature of the crack,$r_{0}$ the interatomic spacing at which the stress becomes zero and $a$ denotes the length of the crack. A simple analysis shows that $\sigma_{C}$ has the dimensions of pressure,which means that the stress multiplied by a volume has the dimensions of energy.

This means that when a fracture forms in a target as a result of the impact the energy ballance has the following form:
\begin{equation}
	\frac{1}{2} m v^{2}-\sigma_{C} V = C_{V} V (T_{1}-T_{0})
\end{equation}
It follows  that in this case the target heats up to a temperature given by
\begin{equation}
	T_{1}=T_{0}+ \frac{1}{C_{V}} \times(\frac{E_{k}}{V}-\sigma_{C})
\end{equation} 
Inserting eqs.(2),(3) and the expression for the volume into eq.(12), simple algebra could give the result for the temperature $T_{1}$ to which the target heats up. 

A physically more interesting approach to the same problem is given by dynamic quantized fracture mechanics ($DQFM$ for short) (for example [5]). The basic difference between $DQFM$ and the usual approach used in material science is that $DQFM$ introduces geometry in studies of scaling laws in material science (for example [1]). Considering that the occurence of a fracture in a material is a sign of its failure, it can be shown in $DQFM$ that the stress necessary for the occurence of a failure is given by ([5]):
\begin{equation}
	\sigma_{f} = K_{Ic}[\frac{1+(\frac{\rho_{0}}{2q})}{\pi (l_{0}+(q/2))}]^{1/2}
\end{equation}
where $K_{Ic}$ denotes the fracture toughness,$\rho_{0}$ is the radius of the crack of length $l_{0}$ and $q$ is the length of the so called fracture quantum. In this case the energy balance is determined by eq.(11), with $\sigma_{f}$ instead of $\sigma_{C}$, and the temperature to which the target heats up is given by eq.(12) with the same replacement.

The final result is  
\begin{eqnarray}
	T_{1}=T_{0}+\frac{3\hbar^{3}}{2 k_{B}}(\frac{\partial P}{\partial \rho})^{3/2}\frac{1}{(k_{B}T_{0})^{3}}[(\frac{3E_{k}}{2\pi a b c})- \nonumber\\
	K_{Ic}[\frac{1+(\frac{\rho_{0}}{2q})}{\pi (l_{0}+(q/2))}]^{1/2}]
\end{eqnarray}
In this way we have obtained an expression for the temperature to which a target heats up after the impact in the case in which a fracture forms in it. Does it or does it not melt as a result of the impact can again be determined by comparing $T_{1}$ with the result of Lindemann's criterion, expressed by eq.(7). It turns out that the condition for "non-melting" is given by 
\begin{eqnarray}
	\frac{E_{k}}{V}-\sigma_{f}\leq C_{V}T_{m0}(\frac{\rho_{0}}{\rho})^{2/3}\exp(2\gamma_{0}(1-(\rho_{0}/\rho)))\nonumber\\
\times[1-\frac{T_{0}}{T_{m0}}\times(\frac{\rho_{0}}{\rho})^{-2/3}\exp(-2\gamma_{0}(1-(\rho_{0}/\rho))]	
\end{eqnarray}
 



\section{The influence of the equation of state}

Several equations in the preceeding section contain the derivative $\frac{\partial P}{\partial \rho}$, which implies that their application demands the knowledge of the equation of state $(EOS)$ of the material of the target. Generally speaking,the $EOS$ is an equation of the form $P=f(\rho,T)$ where all the symbols have their standard meanings,and $f$ is some function. The choice of the $EOS$ appropriate for any given material is a complex task. 

As an example, in the following the so called Birch-Murnaghan $EOS$ will be used. This equation has the form [6]
\begin{eqnarray}
	P(V)=\frac{3B_{0}}{2}\left[(\frac{V_{0}}{V})^{7/3}-\frac{V_{0}}{V})^{5/3}\right]\times\nonumber\\
\left\{1+\frac{3}{4}(B_{0}'-4)\left[(\frac{V_{0}}{V})^{2/3}-1\right]\right\}
\end{eqnarray}
where $B_{0}=-V(\frac{\partial P}{\partial V})_{T}$ is the bulk modulus of the material and $B_{0}'=(\frac{\partial B}{\partial P})_{T}$ is its pressure derivative. The symbols $V_{0}$ and $V$ denote the volume of a specimen under consideration at the initial value of the pressure $P_{0}$ and at some arbitrary value $P$. It can be shown that
\begin{eqnarray}
	\frac{\partial P}{\partial \rho}=\frac{B_{0}}{8\rho_{0}^{3}}[27(B_{0}'-4)\rho^{2}+14(14-\nonumber\\
3B_{0}')\rho_{0}\rho(\frac{\rho}{\rho_{0}})^{1/3}+5(3B_{0}'-16)\rho_{0}^{2}(\frac{\rho}{\rho_{0}})^{2/3}]
\end{eqnarray}
and 
\begin{eqnarray}
(\frac{\partial P}{\partial \rho})^{3/2}=\frac{1}{16\sqrt{2}}(\frac{B_{0}}{\rho_{0}^{3}})^{3/2}[27(B_{0}'-4)\rho^{2}+14(14-\nonumber\\
3B_{0}')\rho_{0}\rho(\frac{\rho}{\rho_{0}})^{1/3}+5(3B_{0}'-16)\rho_{0}^{2}(\frac{\rho}{\rho_{0}})^{2/3}]^{3/2}
\end{eqnarray}
The last expression can be simplified by assuming that $B_{0}'=0$. It follows that in this case the temperature to which the target gets heated as a result of an impact and a fracture forms in it is given by
\begin{eqnarray}
T_{1}=T_{0}+\frac{3 \hbar^{3}}{32 k_{B} \sqrt{2}}(\frac{B_{0}}{\rho_{0}^{3}})^{3/2}\nonumber\\
\times[196\rho_{0}\rho(\frac{\rho}{\rho_{0}})^{1/3}-108\rho^{2}-80\rho_{0}^{2}(\frac{\rho}{\rho_{0}})^{2/3}]^{3/2}\nonumber\\
\times\frac{1}{(k_{B}T_{0})^{3}}\times[\frac{3E_{k}}{2\pi a b c}-K_{Ic}[\frac{1+(\frac{\rho_{0}}{2q})}{\pi (l_{0}+(q/2))}]^{1/2}]
\end{eqnarray}
A similar result could be obtained for any other form of the $EOS$. 

\section{Discussion and conclusions}

The calculations discussed in this paper are mathematically simple,but physically they have considerable interest for planetary science. The main results,expressing the temperature to which a massive target heats up as a result of the impact of a projectile of smaller mass are given by eqs.(6) and (16). The calculations have been performed for two distinct cases: when the impact has no consequences on the structure of the target,and when it is so strong that a fracture occurs in the body of the target. 

There is a distinct difference in the behaviour of the results obtained in the two cases. In what we have called "the ideal case", the temperature of the target always changes as a result of the impact,as $\partial P/\partial \rho$ is always different from zero. No heating effect would occur only in the special case $\partial P/\partial \rho = 0$ which is physically unrealistic, because a material in which pressure and density are not related in some way does not exist.  

An interesting result has been obtained in the "real case". Namely, using stadard material science would have given a result which would have taken into account only the parameters of the material of the target. Instead of such an approach, we have used the $DQFM$ approach, which takes into consideration the material parameters, but also the geometry of the problem. Note that the result contains the length and radius of the crack,as well as the length of the fracture quantum. 

A possible extension of the calculations discussed here could be the inclusion of the presence of magnetic fields. Recent experiments in material science have shown that the presence of low magnetic fields can lead to strenghtening of materials,[4]. This means that their failure strength is increased.   

In this case the heating effect does not always occur. It is clear that for the particular value of the kinetic energy of the impactor, given by
\begin{equation}
	E_{k}= \sigma_{f} V 
\end{equation}
$T_{1}=T_{0}$ in eq.(12),which means that in that case the target does not heat up as a consequence of the impact. Such a result may seem strange, but it is a consequence of the partition of the kinetic energy of the impactor on the heating of the target and formation of a fracture in it.

Results of this paper have obvious applications in planetary science. It is known that impacts and collisions between various bodies have been occuring throughout the history of our planetary system. One of the consequences of these events are the impact craters on the Earth,Moon, Mars,(at least) some asteroids (for example on Eros,see [8]) and satellites. It is expectable that the spots on the surfaces of these objects in which impacts occur get heated. The question which arises is whether or not such events can be analyzed by using solid state physics. Namely, if the material of the target(s) heats up but {\it does not} melt or vaporize, an impact can be analyzed by using laws of solid state physics. However,if melting occurs,solid state physics is applicable only up to the temperature of melting. Details will be discussed elsewhere.     
\section{Acknowledgement}

This paper was prepared within the research project $174031$ financed by the Ministry of Education and Science of Serbia. I am grateful to the referee for helpful comments about the first version of this manuscript.

}


{









{}
\end{document}